\documentclass[12pt]{article}
\usepackage{amsmath}
\usepackage{amsfonts}
\usepackage{graphicx}
\usepackage{xcolor}
\graphicspath{ {../images/} }
\usepackage[margin=1in,top=1.5in,bottom=1.5in]{geometry}

\begin{document}

\title{Critical Dynamics of Random Surfaces\\
 and Multifractal Scaling}

\author{\\[0pt] Christof Schmidhuber\footnote{christof@schmidhuber.ch}\\ [15pt]
Zurich University of Applied Sciences, Switzerland\\ [5pt]
\\  [20pt]
}

\maketitle

\begin{abstract}\vspace{3mm}
The critical dynamics of conformal field theories on random surfaces is investigated beyond the previously studied dynamics of the overall area and the genus.
It is found that the evolution of the order parameter in physical time performs a generalization of the multifractal random walk.
Accordingly, the higher moments of time variations of the order parameter exhibit multifractal scaling.
The series of Hurst exponents is computed and illustrated at the examples of the Ising-, 3-state-Potts-, and general minimal models as well as $c=1$ models on a random surface. 
It is noted that some of these models can replicate the observed multifractal scaling in financial markets.

\end{abstract}

\newpage

\section{Introduction and Summary}

Conformal field theories with central charge $c\le1$ on random surfaces have been proposed as toy models of string theory in two or fewer embedding dimensions.
Their field theory has been established in \cite{poly1,kpz,DDK,seiberg}, while a dual formulation in terms of random matrices has been developed in \cite{brezin,matrix,kaza,mxm,migdal,kleb,david}. 
Recently, it has been argued that these models may also have another application outside of string theory, namely as
the continuum limit of certain social networks that are self-driven to a critical point \cite{me3}. \\

Specifically, it has been proposed to model financial markets
by a Van-der-Waals gas on a lattice, where the lattice represents the social network of traders, while the gas molecules represent the shares of an asset that are distributed across this network \cite{schmi2}.
In this model, {\it efficient} markets correspond to the critical point, where the lattice gas on the network is described by a continuum field theory.
If the network was naively modelled as a static $D$-dimensional hypercubic lattice, this continuum limit would be $\phi^4$ theory,
and choosing $D\approx3$ would replicate the empirically observed second Hurst exponent \cite{schmi2}.\\

In \cite{me3}, a more realistic network model as a {\it random lattice} has been proposed. The Feynman diagrams of the $N\times N$ random matrix models \cite{brezin}
were regarded as two-dimensional small-world networks with a probability $1/N^2$ of long-range links.
Their continuum limit is described by a field theory on a random surface.
To investigate its evolution in time, we must thus develop the critical dynamics \cite{hohenberg} of conformal field theories with $c\le1$ on random surfaces (see \cite{tauber, ZJ} for reviews of critical dynamics).
In this paper, we will pursue this independently of the potential application to financial markets or social networks, 
which merely serves as a motivation, and for which we refer to \cite{me3,schmi2}.
\\

\begin{figure}[t!]\centering
\includegraphics[height=5cm]{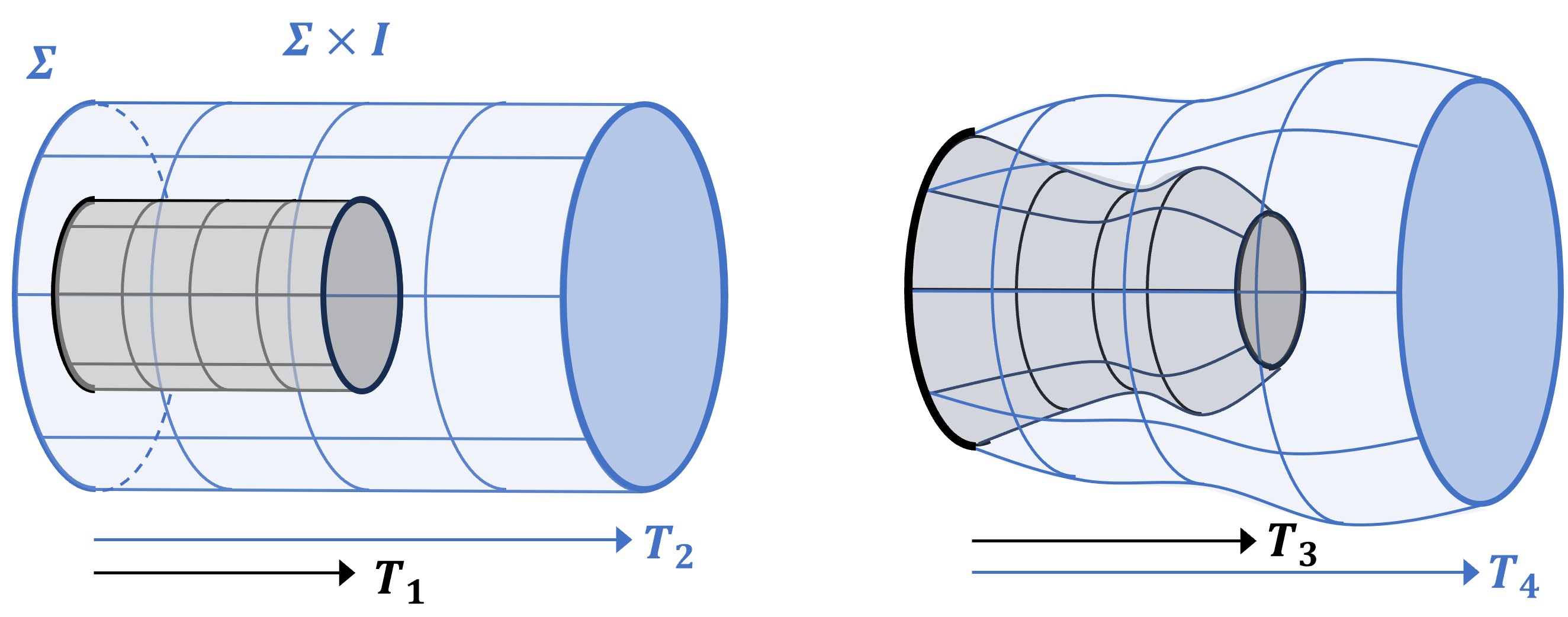}
\caption{In the critical dynamics of random surfaces, the surface $\Sigma$ (drawn as a circle) evolves over a time interval $I$. 
Left: $\Sigma\times I$ in the physical metric $g=\hat g e^\phi$ for $\phi=0$ (gray) and for constant $\phi>0$ (blue). 
Right: two physical metrics of different $\phi_0(t)$ but same $\phi_0(0)$. If background time $\hat T$ is fixed, physical time $T_1, ..., T_4$ is a random variable, and vice versa.} 
\end{figure}

We thus add a third, nonrelativistic time dimension $\hat t$ to the random surface $\Sigma$, which is parametrized by coordinates $\hat\sigma_1,\hat\sigma_2$.
For the dynamics near the critical point, we choose the purely dissipative "model A" of \cite{hohenberg},
which represents the unique dynamic universality class in the absence of conserved quantities.
The two-dimensional conformal symmetry at the critical point then extends to a 2+1-dimensional one under rescalings in space and time:
\begin{equation}
\hat\sigma\rightarrow\lambda\cdot\hat\sigma,\ \ \ \hat t\rightarrow\lambda^z\cdot\hat t,\ \ \ \text{where}\ \ \ \hat\sigma\equiv(\hat\sigma_1,\hat\sigma_2),
\label{skala}\end{equation}
and $z$  is the so-called dynamic critical exponent ($z=2$ for a free field theory).
This makes it convenient to work in a 2+1-dimensional extension of conformal gauge on $\Sigma\times I$, 
where $I$ represents a time interval whose length $\hat T$ is a modulus that must be integrated over. 
Physical distances in space and time are measured in the physical metric $g$:
\begin{equation}
ds^2 = g_{\alpha\beta}\ d\sigma^\alpha d\sigma^\beta,\ \ dt^2=g_{tt}\ d\hat t^2,\ \  \ \text{where}
\ \ \ g_{\alpha\beta}=e^{\alpha\phi(\hat\sigma,\hat t)}\ \hat g_{\alpha\beta},\ \ g_{tt}=e^{{z\over2}\alpha\phi(\hat\sigma,\hat t)}\ \hat g_{tt}.
\label{metric}\end{equation}
Here, $\hat g$ is an arbitrarily chosen background metric that must drop out of physical results, and $\alpha$ is a renormalization parameter.
This should not be confused with a theory of three-dimensional gravity: as in two dimensions, the metric still has only one dynamic component, namely the conformal factor $e^{\alpha\phi(\hat\sigma,\hat t)}$.
We will split $\phi(\hat\sigma,\hat t)$ into three components:
\begin{itemize}\addtolength{\itemsep}{-5 pt} 
\item[(i)] its constant part $\phi_{00}$,
\item[(ii)] its zero-mode, i.e., its integral over the surface at a given time: $\phi_0(\hat t)=\int_\Sigma \phi(\hat\sigma,\hat t)$,
\item[(iii)]  its remaining nonzero modes $\tilde\phi(\hat\sigma,\hat t)$.
\end{itemize}
Fig. 1 (left) shows $\Sigma\times I$ in the physical metric for $\phi=\phi_{00}=0$ (gray) and for $\phi_{00}>0$ (blue);
for clarity, only a cross-sectional circle of $\Sigma$ is drawn. Fig. 1 (right) shows two different physical metrics in the case (ii).
Of course, we must distinguish between the length $\hat T$ of the interval $I$ as measured in background time ($\hat T=4$ in all examples of fig. 1),
vs. its length $T$ as measured in physical time. If $\hat T$ is held fixed, $T$ is a random variable, and vice versa. 
What we are interested in are physical quantities, measured in physical time $T$.
\\

In \cite{me3}, we have begun to study the critical dynamics of random surfaces in an "extended minisuperspace approximation", 
where only the zero mode $\phi_0(\hat t)$ and the genus of the random surface are dynamic variables.
For the dynamics of the zero mode, we have found that the area $A\sim e^{\alpha\phi_0}$ performs a Cox-Ingersol-Ross process in physical time. 
Genus-zero surfaces shrink; to prevent them from shrinking to zero, a small-area cutoff must be imposed. Higher-genus surfaces grow
until their area is of the order of the inverse cosmological constant.\\

Regarding the dynamics of the genus, we have concluded from the matrix model results \cite{mxm,migdal,kleb,david} that it can lead to two distinct phases:
\begin{itemize}\addtolength{\itemsep}{0 pt} 
\item
A {\it planar phase}, in which the ensemble of random surfaces is dominated by surfaces of zero or low genus.
In this phase,  long-range links are irrelevant, and we expect nontrivial critical phenomena as in conformal field theories on a random surface.
\item
A {\it foamy phase}, in which handles condense and all nodes are highly connected, as in a small-world network.
We expect mean field theory to be exact in this phase, as the network effectively reduces to a point in the sense that all distances are of cutoff size.
\end{itemize} 

In the current paper, we investigate the more interesting planar phase in more detail. Specifically, 
we will study the dynamics of an order parameter, such as the magnetization in the case of the Ising model.
%This will take us beyound the minisupespace approximation.
Let $\Phi(\hat \sigma,\hat t)$ be a primary field of dimension $\Delta$ of the conformal field theory that lives on the random surface $\Sigma$, 
such as the spin field with dimension $\Delta=1/8$ in the Ising model.
Our order parameter $\pi(\hat t)$ is then the integral of $\Phi$ over $\Sigma$ at time $\hat t$:
\begin{equation}
\hat\pi(\hat t)=\int_\Sigma d^2\sigma\ \Phi(\hat \sigma,\hat t) .\label{order}
\end{equation}
We consider the moments $M_n(\hat T)$ of the distribution of what we call "returns" of the order parameter, i.e., its
 time variations over a given time horizon $\hat T$:
\begin{eqnarray}
M_n(\hat T)&=&\langle\big[\hat\pi(\hat t+\hat T)-\hat \pi(\hat t)\big]^n\rangle.\label{mom}
\end{eqnarray}
Without gravity (i.e., if $\Sigma$ is a static surface), it is well-known (see, e.g., \cite{ZJ}) that the renormalization group implies that the $n$-th moment scales with $\hat T$ as
\begin{equation}
M_n(\hat T)\ \ \sim\ \ {\hat T}^{nH_n}\ \ \ \text{with}\ \ H_n={1\over z}(1-\Delta),
\label{H0}\end{equation}
where $z$ is the dynamic critical exponent, and the $H_n$ are the so-called Hurst exponents. 
The fact that they are independent of $n$ is called "mono-scaling".\\

How are these Hurst exponents modified "in the presence of gravity", i.e., if $\Sigma$ is a {\it random} surface?
Focusing on small time horizons $T$, we find that the Hurst exponents get "gravitationally dressed" due to short-distance singularities of the correlation functions of the nonzero-modes $\tilde\phi$ (item (iii)).
Here, we can neglect the zero-mode $\phi_0$ (item (ii)), whose correlations are regular at short distances.
The main result of the current paper is the following. If $\Sigma$ is a random surface,
the Hurst exponents exhibit multifractal scaling \cite{mandel} in {\it physical} time in the sense that $H_n$ depends on $n$. 
In the approximation $z\approx2$, we find:
\begin{equation}
H_n\ \ =\ \ {1\over2}(1-\tilde\Delta+\nu_n).\label{hur}
\end{equation}
Here, $\tilde\Delta$ is the well-known KPZ dimension \cite{kpz}
of the primary field $\Phi$ on a two-dimensional random surface, which satisfies
\begin{equation}
\tilde\Delta +{\alpha^2\over4}\tilde\Delta(\tilde\Delta-2)=\Delta,\ \ \ \text{where}\ \ \alpha=\sqrt{25-c\over12}-\sqrt{1-c\over12}. \label{dim}
\end{equation}
The $n$-dependent component $\nu_n$ in (\ref{hur}) is a curious new aspect of the dynamic extension of the theory of random surfaces. We find that it depends only on the central charge $c$:
\begin{equation}
\nu_n=1-{Q\over n\alpha}\Big(1-\sqrt{1-{4n\over Q^2}}\Big),\ \ \ \text{where}\ \ \ Q=\sqrt{25-c\over3}.\label{cor}
\end{equation}
KPZ scaling (\ref{dim}) on a random surface was originally derived in light-cone gauge \cite{kpz}, later rederived in conformal gauge \cite{DDK},
and more recently proven mathematically \cite{duplantier,vargas,davidKPZ}. 
In the dynamic extension, KPZ scaling still describes the scaling of the moments $M_n$ under {\it simultaneous} global rescalings of 
space ($A\rightarrow\lambda^2 A$) and time ($T\rightarrow\lambda^2 T$) as in fig. 1 (left).
However, our results generalize this to {\it independent} rescalings of space and time:
\begin{equation}
M_n\ \sim\ A^{{n\over2}(1-\nu_n)}\cdot T^{{n\over2}(1-\tilde\Delta+\nu_n)}. 
\label{areatime}\end{equation}

For the unitary field theories with $1/2\le c\le1$ on a random surface, only the first two Hurst exponents in (\ref{hur},\ref{dim},\ref{cor}) are real.
As we will see, the distribution of returns has power-law tails that make the higher moments diverge. 
On the other hand, in the limit of large negative central charge, which can occur for non-unitary minimal models, we find that  
the order parameter reduces to a multifractal random walk (MRW) as introduced in \cite{bacry}. In this limit, the higher-order Hurst exponents (\ref{hur},\ref{dim},\ref{cor}) exist and decrease linearly with $n$: 
\begin{equation}
H_n\ \ \approx\ \ {1\over2}(2-\tilde\Delta)-{1\over Q\alpha}-{2n\over Q^3\alpha}.
\label{climit}\end{equation}

This paper is organized as follows. Section 2 briefly reviews the usual mono-scaling for critical dynamics on a flat surface, as well as aspects of random surfaces in conformal gauge.
Section 3 derives correlation functions of the order parameter in the dynamic theory and compares with the multifractal random walk of \cite{bacry}.  
This leads us to multifractal scaling in {\it background} time on a random surface of {\it fixed} total area. However, it
reduces to trivial mono-scaling in background time, once we integrate over the area, i.e., over $\phi_{00}$ (item (i)). \\

Section 4 then takes the key step of translating multifractal scaling in background time (for fixed area) into multifractal scaling in {\it physical} time.
The Hurst exponents $H_n$ are first derived in field theory. They are then rederived by a different method based on stochastic processes, which yields exactly the same formlas (\ref{hur},\ref{dim},\ref{cor}) and explains them geometrically.
Section 5 illustrates the results at the examples of the  Ising model, the 3-state Potts model, other unitary and non-unitary minimal models, and $c=1$ models on a random surface.\\

For certain non-unitary minimal models, these $H_n$ are in fact consistent with the multifractal scaling that has been empirically 
observed in financial markets (see \cite{breymann} for early observations and \cite{bouch2, tizi} for reviews).
This is the subject of a separate paper.

\section{Brief Reviews}

This section briefly summarizes some essential background on conformal field theories with central charge $c\le1$,
their critical dynamics on a static surface, and on random surfaces in conformal gauge. For more extensive reviews, we refer to the literature.

\subsection{Critical Dynamics on a Static Surface}

We first consider two-dimensional conformal field theories (CFTs) on a {\it static} Riemann surface $\Sigma$ of area $\hat A$ (for reviews, see \cite{difrancesco,ginsparg}).
If we want to later put them on a {\it random} surface, it is well-known that we are restricted to CFT's with central charge $c\le1$.\\

For $c<1$, these CFT's are the minimal models \cite{BPZ}.
They include the Ising model, the 3-state Potts model, and many others (see section 5). 
For a given minimal model, there is a set of primary fields $\Phi_{i}, i\in\{1,2,3,...\}$ with dimensions $\Delta_i$.
Some of these models have a Landau Ginzburg description. E.g., the Ising model corresponds to the renormalization group fixed point at $g=g^*$
of scalar $\Phi^4$ theory, the dimension of $\Phi$ being $\Delta=1/8$:
\begin{equation}
S[\Phi]=\int d^2\sigma\Big\{ {1\over2} (\partial\Phi)^2 + V(\Phi)\Big\}\ \ \ \text{with}\ \ \ V(\Phi)=g\cdot\Phi^4.
\label{phi4}\end{equation}
The models with $c=1$ can be represented by a free scalar field on a line, circle, or orbifold.
Some of these models have $O(2)$ or larger continuous symmetry groups. 
For instance, for certain radii of the circle or orbifold, the $c=1$ models are dual to $SO(n)_1\times SO(n)_1/SO(n)_2$ gauged Wess-Zumino-Witten models \cite{witten}, which have global $SO(n)$ symmetries.\\

We now assume that there is an additional time dimension in which these fields evolve, 
in analogy with the time evolution of water and steam at its critical point (which is described by 3-dimensional $\phi^4$ theory).
We would like to compute correlation functions of the order parameter (\ref{order}) at different points in time. This
leads us into the field of critical dynamics \cite{hohenberg}. For in-depth introductions to critical dynamics, see \cite{tauber, ZJ}. A very brief summary in our context is contained in \cite{me3}.\

Given a conformal field theory, we must first choose the dynamics that should govern its time evolution.
It falls into one of several "dynamic universality classes", which differ from each other by the quantities that are conserved.
Without conserved quantities, there is a unique dynamic universality class, the purely dissipative "model A" of \cite{hohenberg}.
In cases where there is a Landau Ginzburg description as in (\ref{phi4}), the equation of motion is
$$\dot \Phi=-{\Omega\over2}{\delta S[\Phi]\over\delta\Phi}={\Omega\over2}\big[\Box \Phi-V'(\Phi)\big],$$
where $\Omega$ has classical dimension 2 and plays the role of Planck's constant.
One can show that its form is preserved by renormalization, except that the dimension of $\Omega$ is corrected to $z$ \cite{hohenberg,tauber,ZJ}.
Here, $z$ is the so-called dynamic critical exponent, which defines the symmetry under extended 2+1-dimensional scale transformations (\ref{skala}).
At 1-loop level, it is related to $\Delta$ by $z=2+c\cdot2\Delta$, where $c\approx 0.7$.  \\

If we want to preserve the order parameter (such as (\ref{order})), we must choose a different universality class called "model B" \cite{hohenberg}.
For the $c=1$ models with $SO(n)$ symmetries, we could choose yet another class ("model J") that conserves the $SO(n)$ currents \cite{hohenberg}.\\

In this paper, we will focus on "model A". We define the order parameter as in (\ref{order}), using a primary field of the CFT, 
and write the moments (\ref{mom}) as integrals:
\begin{eqnarray}
M_n(\hat T)&=&\langle\big[\hat\pi(\hat t+\hat T)-\hat \pi(\hat t)\big]^n\rangle\ \ =\ \ \int_{0}^{\hat T} d^n\hat\zeta\ \langle  \dot{\hat \pi}(\hat\zeta_1) ... \dot{\hat \pi}(\hat\zeta_n) \rangle.\label{fixed}
\end{eqnarray}
The second moment is the variance of time variations of the order parameter $\hat\pi(\hat t+\hat T)-\hat \pi(\hat t)$, which we call "returns". On a static surface $\Sigma$ of area $\hat A$, it scales as \cite{schmi2}
\begin{eqnarray}
M_2&\sim&\hat A\cdot \hat T^{{2\over z}(1-\Delta)}\cdot g_2({\hat T \hat A^{-z/2}}),\ \ \ \text{with}\ \ g_2(x)\rightarrow1\ \ \text{as}\ \ x\rightarrow0,\label{varianz}
\end{eqnarray}
The prefactor $\hat A$ in (\ref{varianz}) reflects translation invariance on the surface $\Sigma$. 
(\ref{varianz}) follows from the renormalization group by requiring the correct behavior under scale transformations in background space and time, 
$$\sigma\rightarrow\lambda\cdot\sigma,\ \hat A\rightarrow\lambda^2\cdot \hat A,\  \hat T\rightarrow\lambda^z\cdot \hat T,\ \hat\pi\rightarrow\lambda^{2-\Delta}\hat\pi
\ \ \Rightarrow\ \ M_2\rightarrow\lambda^{4-2\Delta}M_2,$$ 
as well as consistency with the limit $\Delta=0, z=2$ of an ordinary random walk with linearly growing variance $M_2\sim \hat T$.
In (\ref{varianz}), we have included an arbitrary function $g_2(x)$ of the scale-invariant combination $x=\hat T \hat A^{-z/2}$, 
which is allowed by the renormalization group but must drop out in the limit $\hat A\rightarrow\infty$ or $\hat T\rightarrow0$. For higher moments, scaling implies:
\begin{eqnarray}
M_n(\hat T)&\sim&\hat A^{n\over2}\cdot \hat T^{{n\over z}(1-\Delta)}\cdot g_n({\hat T \hat A^{-z/2}}),\ \ \ \text{with}\ \ \ g_n(x)\rightarrow1\ \ \text{as}\ \ x\rightarrow0.\label{mono}
\end{eqnarray}
As remarked in the introduction, for short time horizons $\hat T$ we obtain the "mono-scaling" (\ref{H0}) with equal Hurst exponents.

\subsection{Random Surfaces in Conformal Gauge}

Let us now put our 2-dimensional CFT with primary fields $\Phi_i$, $i\in\{1,2,...\}$, of scaling dimensions $\Delta_i$ 
on a random surface of genus $0$. In conformal gauge, we can locally
write the two-dimensional metric as $g_{ab}=\hat g_{ab}e^{\phi}$, where $\hat g_{ab}$ is
an arbitrarily chosen background metric with curvature $\hat R_{ab}$, and $\phi$ is the logarithm of the conformal factor.
Recall that the following effective action for $\phi$ arises from the conformal anomaly \cite{poly1,kpz,DDK,seiberg}:
\begin{eqnarray}
S&=&{1\over8\pi}\int d^2\hat\sigma\sqrt{\hat g}\ \hat g^{ab}\{\partial_a \phi\partial_b\phi +Q\hat R_{ab}\phi+\mu e^{\alpha\phi}\}+\beta\int d^2\hat\sigma\sqrt{\hat g}\ \Phi_i e^{\alpha_i\phi},\label{anomaly}
\end{eqnarray}
where $\alpha, \alpha_i,$ and $Q$ are renormalization parameters, $A=\int d^2\sigma\sqrt{\hat g}\ e^{\alpha\phi}$ is the physical area, 
$\mu$ is the two-dimensional cosmological constant, $\beta$ is a small coupling constant, 
and $e^{\alpha_i\phi}$ is the so-called ``gravitational dressing" of $\Phi_i$.
This theory must be independent of the arbitrary background metric $\hat g$, and, in particular, scale invariant.
Therefore the central charge of the combined matter-$\phi$ theory must be zero,
and the operators $e^{\alpha\phi}$ and $\Phi_i e^{\alpha_i\phi}$ must have dimension 2 before integration over the surface. 
This can been seen to imply \cite{DDK,seiberg}:
\begin{eqnarray}
Q^2={1\over3}(25-c)\ \ \ ,\ \ \ \alpha(Q-\alpha)=2\ \ \ ,\ \ \ \alpha_i(Q-\alpha_i)=2-\Delta_i,\label{DDK}
\end{eqnarray}
where we take the negative branches of the solutions for $\alpha$ and $\alpha_i$, and $Q>0$. 
E.g., for the Ising model, one obtains $\alpha^2=3/2,\ {Q/\alpha}={7/3}$, and  ${\alpha_i/\alpha}={5/6}$.\\

What is the dimension $\tilde\Delta_i$ of the field $\Phi_i$ {\it after} it has been put on the random surface? 
This question may seem puzzling at first, as the theory must be completely scale invariant with respect to the background metric.
Then how can there be a nontrivial dimension?
The answer is, we must examine the behavior of $\beta e^{\alpha_i\phi}$ under {\it physical}, not {\it background} scale transformations.
Since the area $A=\int e^{\alpha\phi}$ has dimension $-2$, 
{\it physical} (as opposed to {\it background}) rescalings by a factor $\lambda$ correspond to constant shifts of the field $\phi$: 
\begin{eqnarray}
\phi\rightarrow\phi+{2\over\alpha}\ln\lambda&\Rightarrow& 
A\rightarrow \lambda^{2}A,\ \ \ \int \Phi_{i}e^{\alpha_{i}\phi}\rightarrow \lambda^{2-\tilde\Delta_i}\int\Phi_{i}e^{\alpha_{i}\phi}\notag\\
&&\text{where}\ \ \ \tilde\Delta_i=2-2{\alpha_i\over\alpha}.\label{shift}
\end{eqnarray}
Thus, the physical scale dependence is encoded in the 
$\phi$-dependence of $\beta e^{\alpha_i\phi}$. Therefore, the ``gravitationally dressed" dimension of $\Phi_i$ is 
$\tilde\Delta_i$ before integrating over the surface. E.g., for the Ising model, $\tilde\Delta_i=1/3$.
It is easy to see that $\tilde\Delta_i$ satisfies the KPZ relation (\ref{dim}).

\subsection{Correlation Functions of the Conformal Factor}

In the presence of gravity, i.e., on a random surface, the field $\Phi_i$ in (\ref{order}) and thereby the order parameter gets "gravitationally dressed" as shown in (\ref{anomaly}):
$$\int_\Sigma d^2\sigma\sqrt{\hat g}\ \Phi_i(\sigma)\ \ \ \rightarrow\ \ \ \int_\Sigma d^2\sigma\sqrt{\hat g}\ \Phi_i(\sigma)\ e^{\alpha_i\phi(\sigma)}.$$
The dressed versions of the moments (\ref{fixed}) will thus involve correlation functions of the dressing operators $e^{\alpha_i\phi}$.
Let us first recall their correlations without time dimension.
We choose a constant curvature background $\hat g_{ab}$ and  
split the conformal factor into the spatially constant mode $\phi_{00}$ and the remainder $\tilde\phi(\hat\sigma)$:
$$\phi_{00}=\int_\Sigma d^2\sigma\ \phi(\sigma)\ ,\ \ \tilde\phi(\sigma)=\phi(\sigma)-\phi_{00}.$$
For $\mu=\beta=0$ in (\ref{anomaly}), only $\phi_{00}$ sees the so-called background charge Q; its action on a genus-zero surface (whose Euler characteristic is 2) is simply $Q\phi_{00}$.
The equal-time propagator of the nonzero modes $\tilde\phi$, treated as a free field, is well-known:
\begin{eqnarray}
G^{(2)}(\sigma_1-\sigma_2)&\equiv&\langle\tilde\phi(\sigma_1)\tilde\phi(\sigma_2)\rangle\ =\ -\ln \vert \sigma_1-\sigma_2\vert^2.\notag
\end{eqnarray}
From this, the equal-time correlation functions of the gravitational dressing operators are obtained by Wick contractions, using
 $\langle e^{\tilde\phi(\sigma_1)} e^{\tilde\phi(\sigma_2)}\rangle=e^{\langle\tilde\phi(\sigma_1)\tilde\phi(\sigma_2)\rangle}$ (see, e.g., \cite{difrancesco}):
\begin{eqnarray}
 \langle\prod_{i=1}^n e^{\alpha_{i}\tilde\phi(\sigma_i)}\rangle&=&\prod_{k<l}\vert \sigma_k-\sigma_l\vert^{-2\alpha_{i}^2}\ \sim\ \lambda^{{1\over2}n(n-1)\alpha_i^2}\ \ \ \text{under}\ \ \ \sigma\rightarrow\lambda\cdot\sigma. \label{2dscal}
\end{eqnarray}
However, this scaling holds {\it only} if the expectation value does {\it not} include an integral over the zero mode $\phi_{00}$,
i.e., if the "zero mode area" $A=e^{\alpha\phi_{00}}$ is held fixed.
Indeed, the zero mode integral with action (\ref{anomaly}) at $\beta=0$ can be evaluated (see, e.g., \cite{schomerus} for a review):
\begin{eqnarray}
\int_{-\infty}^{+\infty} d\phi_{00}\ \exp\{-s\cdot\alpha\phi_{00}-e^{\alpha\phi_{00}}\cdot X\} &\sim&\Gamma(s)\cdot X^s \notag\\
\text{with}\ \ \ s\equiv{Q\over\alpha}-\sum_{i=1}^n{\alpha_i\over\alpha},&&X\equiv {\mu\over8\pi}\int_\Sigma e^{\alpha\tilde\phi(\hat\sigma)}.
\end{eqnarray}
We see that integrating over $\phi_{00}$ inserts $s$ cosmological constant operators $X$ on the left-hand side of (\ref{2dscal}),
where we use the "replica trick" of evaluating the integral for integer $s$ and then analytically continuing to real $s$. 
As a result ``Liouville charge is conserved" in (\ref{2dscal}) in the sense that $n\alpha_{i}+s\alpha=Q$.
This in turn leads to a scaling exponent that is linear in $n$ under rescalings of background time (using (\ref{DDK})):
\begin{eqnarray}
\langle\prod_{k=1}^n e^{\alpha_{i}\tilde\phi(t_k)}\rangle&\sim&\lambda^{-{n\over2}\alpha_{i}(Q-\alpha_{i})}\ \sim\ \lambda^{{n\over2}(\Delta_i-2)}.
\end{eqnarray}
This is just what is needed to ensure that the dressed order parameter in (\ref{anomaly}) is invariant under background scale transformations,
as it must be, after accounting for the scaling dimension $\Delta_i$ of $\Phi_i$ and the integral over $\Sigma$.

\section{Critical Dynamics of the Order Parameter}

We would now like to generalize the scaling (\ref{mono}) to the case where the CFT lives on a random surface.
Appart from the correlation functions (\ref{fixed}) of the CFT itself, this will involve correlations of the field $\phi$,
which we have split into three components in the introduction: (i) the constant mode $\phi_{00}$,
(ii) the zero mode $\phi_0(\hat t)$, and (iii) the nonzero modes $\tilde\phi(\hat\sigma,\hat t)$.  \\

The zero mode (ii) has been discussed extensively in \cite{me3}.
Its correlation functions do not contribute to the singularities at small $\hat t$, 
and thus do not affect the scaling at short time scales, so we will ignore it here.
This leaves us with (i) and (iii). 
To simplify the calculation, we now approximate $z=2+c\cdot\Delta_i\approx2$, which is reasonable for $\Delta_i\ll1$.

\subsection{Scaling Ansatz in Physical Time}

The constant mode (i) generates simultaneous global scale transformations (\ref{skala}) in space and time.
A natural ansatz for the analog of the scaling (\ref{mono}) on a random surface in the case $z=2$ involves the gravitationally dressed (KPZ) dimension $\tilde\Delta_i$ in (\ref{shift}):
\begin{eqnarray}
M_n(T,t)\propto A^{n\over2}\cdot T^{{n\over 2}(1-\tilde\Delta_i)}\cdot f_n(TA^{-1})\label{ansatz}
\end{eqnarray}
where  $f_n$ is some analytic function. $A$ and $T$ now denote the {\it physical} area and the {\it physical} time,
as measured in the physical metric $g_{ij}$ in (\ref{metric}), as opposed to the background area $\hat A$ and the background time $\hat T$, which are measured in the background metric $\hat g_{ij}$.
$A$ in (\ref{ansatz}) can be chosen as either the initial area, or some weighted average area over the time interval $T$.
We find it convenient to define it as the ``zero-mode area" $A\equiv e^{\alpha\phi_{00}}$.\\

(\ref{ansatz}) satisfies global physical scale invariance, i.e., invariance under constant shifts of $\phi$:
$$\phi\rightarrow\phi+{2\over\alpha}\ln\lambda\ \ \ \Rightarrow\ \ \ A\rightarrow \lambda^{2}A,\ \ \ T\rightarrow \lambda^{2}T,\ \ \ M_n\rightarrow \lambda^{2n\alpha_i/\alpha}M_n.$$
However, global scale invariance does not determine the functions $f_n(x)$ of the scale invariant combination $x=TA^{-1}$. 
In the next section, we will derive a power law for $f_n$:
\begin{equation}
f_n(x)\rightarrow x^{{1\over2}n\nu_n}\ \ \ \text{as}\ \ \ x\rightarrow0,\label{nu}
\end{equation}
where the exponent $\nu_n\in R$ is given by (\ref{cor})
(recall that $\nu_n=0$ on a static surface according to (\ref{mono})).
After dividing both sides of equation (\ref{ansatz}) by $e^{n\alpha_i\phi_{00}}$, the
constant mode $\phi_{00}$ drops out of the equation.
This leaves us with the task of deriving $f_n(T)$ in (\ref{ansatz}) only from the nonzero modes $\tilde\phi$
on a surface of fixed $\phi_{00}=0$, i.e., of fixed zero mode area $A=1$.\\

The stochastic differential equation for $\tilde\phi$ in "model A" is \cite{ZJ, tauber, me3}:
\begin{equation}
{d\over d\hat t} {\tilde\phi}\ =\ -{1\over2}{\delta S[\tilde\phi]\over\delta\tilde\phi}+\eta\ =\ {1\over2}\Box \tilde\phi-{\mu\over2}\alpha e^{\alpha\phi_{00}} e^{\alpha\tilde\phi(\hat t)}+\eta,
\label{EM}\end{equation}
where the random noise $\eta$ satisfies $\int_\Sigma\eta=0$, and we have set $\Omega=1$ by rescaling background time.
We will study the moments for time scales $T\ll\mu^{-1}$, where we can neglect the cosmological constant.
Then $\tilde\phi$ can be treated like a free field, whose dynamic critical exponent is $z_\phi=2$. 
Note that the condition $\int_\Sigma\tilde\phi=0$ is then automatically preserved, as $\dot {\tilde\phi}$ is a divergence.
Thus, there is no need to switch to the more complex dynamics of "model B" of \cite{hohenberg}, 
which is often used for models with a conserved order parameter. 

\subsection{Dynamic Correlation Functions}

The covariant generalization of the time derivative of the order parameter $\dot{\hat\pi}_i$ in (\ref{fixed}) to a curved 
surface with metric $g_{ab}=\hat g_{ab} e^{\alpha\tilde\phi}$ (but still in background time $\hat t$ with $\hat g_{tt}=1$) is 
\begin{eqnarray}
\dot{\hat\pi}_i(\hat t)\ \rightarrow\ O_i(\hat t)\ &\equiv& \int_\Sigma d^2\sigma\sqrt{\hat g}\ e^{\alpha_{i}\tilde\phi(\sigma,\hat t)}\cdot {\partial\over \partial\hat t}\Phi_i(\sigma,\hat t),  \label{dress}\\
M_n(\hat T)\ =\ \big\langle[\pi_i(\hat T)-\pi_i(0)]^n\big\rangle &=&\int_0^{\hat T} d^n\hat t\ \big\langle O_i(\hat t_1)... O_i(\hat t_n)\big\rangle.\notag
\end{eqnarray}
The moments now also contain correlation functions of the gravitational dressing operators $e^{\alpha_i\tilde\phi}$.
Their correlation functions at equal times, but at different points on the surface, are given by (\ref{2dscal}).
Using $z_\phi=2$, the symmetry under rescalings $\vec x\rightarrow\lambda\cdot\vec x,\ \hat t \rightarrow\lambda^2\cdot t$  then implies the following scaling of 
the correlation functions of the dressed order parameter in time:
\begin{eqnarray}
 \big\langle\prod_{k=1}^n O_i(\hat t_k)\big\rangle&=&\big\langle  \dot{\hat \pi}_i(\hat t_1) ... \dot{\hat \pi}_i(\hat t_n) \big\rangle\cdot C_n
\ \ \ \text{with}\ \ \ C_n\sim\prod_{k<l}\vert \hat t_k-\hat  t_l\vert^{-\alpha_{i}^2}.\label{bacry2}
\end{eqnarray}
Here, we omit powers of $\hat A$, which must be such that $M_n$ is invariant under background scale transformations. 
(\ref{bacry2}) can be derived directly in the path integral formalism \cite{ZJ}: for $\mu=0$, the equation of motion (\ref{EM}) is derived from the free field action 
\begin{equation}
S[\tilde\phi]={1\over2}\int d^2\hat\sigma\ d\hat t\ (\dot{\tilde\phi}-{1\over2}\Box\tilde\phi)^2. 
\label{S}\end{equation}
The free field propagator in two-dimensional momentum- and position space is \cite{tauber, ZJ}: 
$$G^{(2)}(\hat p,\hat t)={1\over \hat p^2}e^{-\hat p^2\hat t/2}\ \Rightarrow\ \partial_{\hat t} G^{(2)}(\hat \sigma,\hat t)
\sim-{1\over \hat t} e^{-{\hat \sigma^2/(2\hat t)}}\ \Rightarrow\ G^{(2)}(\hat \sigma,\hat t)\sim\int_{\hat t}^\tau {d\zeta\over \zeta} e^{-{\hat\sigma^2/(2\zeta)}}$$
with correlation time $\tau$. Here, we have fixed the integration constant by requiring $G^{(2)}=0$ at $\hat t=\tau$.
$M_n$ in (\ref{dress}) has a saddle point at $\hat\sigma_i=\hat\sigma_j$ for all $i\neq j$.
Since $G^{(2)}(0,\hat t)=(\ln\tau-\ln \hat t)$, the saddle point value, as computed by Wick contractions, is indeed given by (\ref{bacry2}).
For small $\hat t$, $M_n$ is dominated by the vicinity of this saddle point.
Let us approximate the heat kernel $\partial_{\hat t}G^{(2)}$ by a uniform distribution with support on a disk $D$ of linearly growing radius $\hat t$: 
\begin{eqnarray}
\partial_t G(\hat \sigma,\hat t)\ \sim\ {1\over \hat t}\cdot\vartheta(\hat t-\hat \sigma^2)\ \ \ \Rightarrow\ \ \  G(\hat \sigma,\hat t)\ \sim\  
\left\{\begin{array}{ll}
        \ln\tau-\ln \hat t & \text{for } \hat \sigma^2\le\hat t \\
        \ln\tau-\ln \hat \sigma^2 &\text{for } \hat \sigma^2>\hat t\\
\end{array}\right. \label{derivatives} 
\end{eqnarray}
The disk $D$ thus contributes (\ref{bacry2}) to $C_n$,
while its complement on $\Sigma$ contributes a less singular term that we can neglect for small $\hat t$. One may verify that evaluating
$G^{(2)}$ exactly using the exponential integral function instead of approximation (\ref{derivatives}) leads to the same conclusion.
\\

We can now re-interpret the $C_n$ in (\ref{bacry2}) as correlation functions
of a new mode, which depends only on time and which we also call $\tilde\phi(\hat t)$, with logarithmic propagator: 
\begin{eqnarray}
C_n(\hat t)&\equiv&\big\langle e^{\alpha_{i}\tilde\phi(\hat t_1)}\ ...\ e^{\alpha_{i}\tilde\phi(\hat t_n)}\big\rangle
\ \ \ \text{with}\ \ \ \big\langle\tilde\phi(0)\tilde\phi(\hat t)\big\rangle= (\ln\tau -\ln \hat t).\label{fix}
\end{eqnarray}
In other words, when computing correlation functions of the order parameter, 
we can effectively replace all the nonzero modes of the 2+1-dimensional field $\tilde\phi(\hat \sigma,\hat t)$ by the single new 1-dimensional mode $\tilde\phi(\hat t)$.
It can be thought of attaching charges $\alpha_{i}$ to the ``particles" $\dot{\hat\pi}_i(\hat t)$ with an attractive logarithmic potential.
(\ref{dress}) thus simplifies to
\begin{eqnarray}
\dot{\hat\pi}_i(\hat t)&\rightarrow& e^{\alpha_{i}\tilde\phi(\hat t)}\dot{\hat\pi}_i(\hat t)\ \ \ \Rightarrow
\ \ \ {\hat\pi}_i\rightarrow \pi_i\equiv \int_0^{\hat T}d\hat t\ e^{\alpha_{i}\tilde\phi(\hat t)}\dot{\hat\pi}_i(\hat t).\label{bac}
\end{eqnarray}

\subsection{Multifractal Random Walk (MRW)}

(\ref{bacry2},\ref{fix},\ref{bac}) precisely replicate the ``multifractal random walk" introduced in \cite{bacry}.
Let us briefly summarize the a few aspects of it.
Viewed as a stochastic process, $\hat\pi(\hat t)$ in (\ref{order}) may be a Gaussian random walk with $\Delta=0$, or a fractional random walk with $\Delta\neq0$.
The authors of \cite{bacry} consider the situation where $\hat\pi(\hat t)$ is coupled to a Gaussian random variable $\gamma\tilde\phi(\hat t)$ as in (\ref{bac}),
with variance $\gamma^2$ and logarithmic covariance in time as in (\ref{fix}): 
\begin{equation}
\big\langle\tilde\phi(0)\tilde\phi(\hat t)\big\rangle=(\ln\tau-\ln\hat t)\ \ \ \Rightarrow  \ \ \ \big\langle\tilde\phi^2\big\rangle=\ln\tau.
\label{bacry}\end{equation}
The divergences at $\hat t\rightarrow0,\ \tau\rightarrow\infty$ are rgularized by a short-time cutoff $\Delta \hat t$
and a correlation time $\tau$. $\gamma$ is a free real parameter of the model,
and $e^{\gamma\tilde\phi(\hat t)}$ is interpreted as ``stochastic volatility" in the sense that it multiplies the time variations of $\hat\pi(\hat t)$ as in (\ref{bac}):
$$\dot{\hat\pi}(\hat t)\rightarrow \dot\pi(\hat t)\equiv e^{\gamma\tilde\phi(\hat t)}\dot{\hat\pi}(\hat t).$$
This leads to the moments (\ref{bacry2},\ref{fix}) with $\gamma$ in the role of $\alpha_i$.
In our model, the field $\tilde\phi$ has a natural origin: it is a remnant of the conformal factor on the random surface $\Sigma$.
However, we are not free to choose $\gamma$, but $\gamma=\alpha_i$ is fixed by the central charge and $\Delta$ through relations (\ref{DDK}). 
Note that, compared with the authors of \cite{bacry}, who work with the bare field $\gamma\tilde\phi_B$, 
we work with the renormalized field $\tilde\phi=\tilde\phi_B-\ln (\Delta \hat t/\tau)$.
This removes the divergent expectation value of $\tilde\phi_B$ of \cite{bacry}, while $e^{\gamma\tilde\phi}$ acquires an anomalous dimension $\gamma^2/2$.\\

The following has already been shown in \cite{bacry}: in
the simpler case of a Gaussian random walk, $\langle\dot{\hat\pi}(\hat t_1)\dot{\hat \pi}(\hat t_2)\rangle=\delta(\hat t_1-\hat t_2)$. 
Integrating out $\hat\pi$ in (\ref{bacry2}) thus pairs the operators $e^{\gamma\tilde\phi}$ into $e^{2\gamma\tilde\phi}$. For even $n$,
there are $n(n-2)/8$ links between such pairs, which yields the scaling
$$M_{n}\propto\int_0^{\hat T} d^{n\over2}\zeta\prod_{i<j}\vert \zeta_i-\zeta_j \vert^{-4\gamma^2}\ \sim\ {\hat T}^{{n\over2}+{n\over2}(2-n)\cdot\gamma^2}.$$
This yields the Hurst exponents
$$H_n={1\over2}+{2-n\over2}\cdot\gamma^2.$$
If $\pi$ is a fractional random walk with $\Delta\neq0$, one instead obtains the Hurst exponents \cite{bacry}
\begin{equation}
H_n={1\over 2}(1-\Delta)+{1-n\over 2}\cdot\gamma^2.\label{hurst3}
\end{equation}
The $H_n$ thus display multifractal scaling with the time interval $\hat T$, decreasing linearly with $n$. This implies that the shape of the return distribution is not scale invariant.
In particular, its tails are "fatter" for shorter time horizons. E.g., the kurtosis $M_4/M_2^2\sim \hat T^{-4\gamma^2}$ decreases with $\hat T$ until it reaches some value $\ge3$ (the Gaussian value) 
at the correlation time. \\

In the context of two-dimensional gravity, a multifractal analysis has been discussed, e.g., in \cite{vargas,LiouQG}.
Note that this differs from our multifractal scaling in time, which applies to the dynamic extension.
Also, as in subsection 2.3, multifractal scaling in background time arises {\it only} if we fix the zero-mode $\phi_{00}$, i.e., if we fix the area. 
Once we integrate over $\phi_{00}$, we must take the cosmological constant into account. As in subsection 2.3, this  would
lead to a scaling exponent that is linear (not quadratic) in $n$ under rescalings of background time,
\begin{eqnarray}
\big\langle\prod_{k=1}^n e^{\alpha_{i}\tilde\phi(\hat t_k)}\big\rangle&\sim&\hat T^{-{n\over2}\alpha_{i}(Q-\alpha_{i})}.
\end{eqnarray}
It would reduce our multi-scaling in background time to simple monoscaling.
This is consistent with the fact that the combined matter-$\phi$ theory is a CFT in the background metric, 
where the dressed order parameter has dimension 2 and (\ref{mono}) should apply. 

\subsection{Background vs. Physical Time}

So far, we have seen that the order parameter of a CFT on a random surface of fixed "zero mode area" $A=e^{\phi_{00}}=1$ performs a multifractal random walk 
with Hurst exponents of the form (\ref{hurst3}) due to its gravitational dressing.
However, this is an MRW in the {\it background} time scale, 
while we are really interested in the evolution (\ref{ansatz}) in the {\it physical} time scale.
We must thus translate the moments 
\begin{eqnarray}
M_n(\hat T)&=& \big\langle m_n(\hat T)\big\rangle\ \ \ \text{with}\ \ \ m_n=\int_0^{\hat T}d^n\hat t\  \prod_{k=1}^n \dot{\hat\pi}_i(\hat t_k)\ e^{\alpha_i\tilde\phi(\hat t_k)}
\label{pmom}\end{eqnarray}
from functions of background time $\hat T$ into functions of physical time 
\begin{equation}
T(\hat T)=\int_0^{\hat T}d\hat t\ e^{\alpha\tilde\phi(\hat t)}.\label{time}
\end{equation}
That is, we must evaluate the expectation value (\ref{pmom}) for all paths $\tilde\phi(\hat t)$, for which the physical time interval (\ref{time}) has some fixed value $\theta$,
thereby integrating over all possible values of the modulus $\hat T$.
Since $\phi_{00}=0$ and therefore $A=1$ is held fixed, the scaling in background time (\ref{bacry2}) applies, as discussed in the previous subsection, and
(\ref{ansatz}) reduces to
\begin{eqnarray}
M_n(T,t)\propto T^{{n\over 2}(1-\tilde\Delta_i)}\cdot f_n(T).\label{ansatz2}
\end{eqnarray}

\section{Hurst Exponents on a Random Surface}

In this section, we derive the $f_n$ in (\ref{ansatz2}) to obtain the scaling in physical time $T$. 
In subsection 4.1, we present a formal computation of the Hurst exponents. 
In subsection 4.2, we confirm and intuitively explain the results by rederiving them in terms of stochastic processes. 

\subsection{Algebraic Derivation}

To perform the path integral over $\tilde\phi(t)$ in (\ref{pmom}), we must rely on the background CFT formulation.
However, in this formulation there is a difficulty in switching from background time $\hat T$ to physical time $T$: 
$T$ is not a real function of $\hat T$, but an operator.
To overcome this difficulty, we use a method employed in \cite{davidKPZ} to rederive KPZ scaling (\ref{dim}) on a random surface without time; we adopt it to our situation, where we also have the time dimension.
Using the operators $m_n$ and $T$ defined in (\ref{pmom},\ref{time}), we consider the analytic function $F_n(\kappa)$, defined for integer $\kappa$ as the correlation function
\begin{equation}
F_n(\kappa)\ \equiv\ \big\langle m_n T^{\kappa}\big\rangle\ =\ \int_0^\infty {d\hat T\over \hat T}\ I_{n,\kappa}(\hat T)
\ \ \ \text{with}\ \ \ I_{n,\kappa}(\hat T)\equiv\big\langle m_n(\hat T)\cdot \big[T(\hat T)\big]^{\kappa}\big\rangle.\label{mellin}
\end{equation}
Here, background time $\hat T$, which is a modulus of conformal gauge, is integrated out using the conformally invariant integration measure $d\hat T/\hat T$.
On the other hand, inserting a delta function that restricts to fixed physical time $\theta$ and then integrating over $\theta$, 
we see that $F_n(\kappa)$ is the Mellin-Barnes transform of the $n$-th moment $M_n(\theta)$:
$$F_n(\kappa)\ =\ \int_0^\infty {d\theta\over\theta}\ \big\langle m_n T^{\kappa}\cdot\delta(\ln T-\ln \theta) \big\rangle\ =\ \int_0^\infty d\theta\ M_n(\theta)\cdot \theta^{\kappa-1}\ .$$
Assuming the scaling $M_n(\theta)\rightarrow \theta^{nH_n}$ for small time intervals ($\theta\rightarrow0$), where $H_n$ is the $n$-th Hurst exponents,
the first pole of the Mellin-Barnes transform $F_n(\kappa)$ occurs at $\kappa=-nH_n$. 
This first pole is easily computed in background time by performing integral (\ref{mellin}). 
Using (\ref{pmom}, \ref{time}), we can write out the integrand in (\ref{mellin}) in the case where $\kappa$ is integer: 
\begin{eqnarray}
I_{n,\kappa}(\hat T)&=& \int_0^{\hat T}d^n\hat t \int_0^{\hat T}d^\kappa\hat s\ \big\langle  \dot{\hat \pi}_i(\hat t_1)\ ...\ \dot{\hat \pi}_i(\hat t_n) \big\rangle
\cdot\big\langle e^{\alpha_{i}\tilde\phi(\hat t_1)}\ ...\ e^{\alpha_{i}\tilde\phi(\hat t_n)}\ \ e^{\alpha\tilde\phi(\hat s_1)}\ ...\ e^{\alpha\tilde\phi(\hat s_\kappa)}\big\rangle.\notag
\end{eqnarray}
There are $n(n-1)/2$ pairs of operators $e^{\alpha_i\tilde\phi}$, $\kappa(\kappa-1)/2$ pairs of operators $e^{\alpha\tilde\phi}$, and $n\kappa$ cross pairs.  
Using  (\ref{mono}) and Wick's theorem with (\ref{fix}), and simplifing the exponent using (\ref{DDK}),  the integrand thus scales for $z\approx2$ and small $\hat T$ as
\begin{eqnarray}
I_{n,\kappa}(\hat T)&\sim& \hat T^{{n\over2}(1-\Delta_i)+\kappa-{1\over2}n(n-1)\alpha_i^2-{1\over2}\kappa(\kappa-1)\alpha^2-n\kappa\alpha\alpha_i}\notag\\
&=& \hat T^{-{n\over2}+{1\over2}Q(n\alpha_i+\kappa\alpha)-{1\over2}(n\alpha_i+\kappa\alpha)^2}.\notag
\end{eqnarray}
Using the "replica trick" as in \cite{davidKPZ}, we analytically continue the results to complex $\kappa$. The pole of the Mellin-Barnes transform $F_n$ occurs when the exponent is zero, i.e., at 
$$(n\alpha_i+\kappa\alpha)^2-Q(n\alpha_i+\kappa\alpha)+n=0.$$
Solving for $\kappa=-nH_n$ and choosing the negative branch, we obtain the Hurst exponents
\begin{equation}
H_n\ =\ {\alpha_i\over\alpha}-{Q\over 2n\alpha}\Big(1-\sqrt{1-{4n\over Q^2}}\Big),\label{HH1}
\end{equation}
or, recalling (\ref{shift}) and defining $\nu_n$ as the deviation from the naive expectation $H_n={1\over2}(1-\tilde\Delta_i)$:
\begin{equation}
H_n\ \ =\ \ {1\over2}(1-\tilde\Delta)+{1\over2}\nu_n\ \ \ \text{with}\ \ \ \nu_n=1-{Q\over n\alpha}\Big(1-\sqrt{1-{4n\over Q^2}}\Big).\label{HH3}
\end{equation}
This is the result stated in the introduction (\ref{hur},\ref{dim},\ref{cor}). Note from (\ref{DDK}) that the multi-scaling correction $\nu_n$ depends only on the central charge,
not on the dimension $\Delta_i$ of the order parameter. Note also that the second Hurst exponent $H_2$ {\it is} the naive one ($\nu_2=0$).\\

The Hurst exponents (\ref{HH1}) are real for $n\le Q^2/4=(25-c)/12$. Since $c\le1$, this shows that the second moment (the variance of returns) exists for all minimal models on a random surface.
However, the third and fourth moments (skewness and kurtosis) exist only for non-unitary models with negative central charge $c\le-11$ and $c\le-23$, respectively. 
This indicates that the tails of the return distributions are too "fat" for the other models.
Note that a linear decrease of $H_n$ with $n$, as for the MRW in (\ref{hurst3}), holds only approximately in the limit of large $Q$, i.e., of large negative central charge. Then, as anticipated in (\ref{climit}),
\begin{equation}
H_n\ \approx\ {\alpha_i\over\alpha}-{1\over Q\alpha}-{2n\over Q^3\alpha}\ \ \ \text{for}\ \ Q\rightarrow\infty .
\label{HH2}\end{equation}

We can now return to our scaling ansatz (\ref{ansatz}). 
Given the Hurst exponents (\ref{HH3}), in order to obtain the correct scaling
of the moments with physical time $T$ we need
\begin{eqnarray}
f_n\Big({T\over A}\Big)\sim \Big({T\over A}\Big)^{{1\over2}n\nu_n}\ \ \ \Rightarrow\ \ \ M_n& \sim& A^{{n\over2}(1-\nu_n)}\cdot T^{{n\over2}(1-\tilde\Delta+\nu_n)} ,\label{ansatz1}
\end{eqnarray}
which confirms (\ref{areatime}). From (\ref{DDK}), $\nu_1>0, \nu_2=0$ and $\nu_n\le0$ for $n>2$ if it is real at all.\\

(\ref{ansatz1}) also extracts the scaling with respect to the physical area $A$. From this, we obtain the overall Hurst exponents (without fixing the area)
by integrating (\ref{ansatz1}) over $A$, weighted by the fixed-area partition function $Z(A)\sim A^{-1-Q/\alpha}e^{-\mu A}$ as discussed in \cite{me3}.
As explained in section 3, this reduces multifractal scaling in {\it background} time to simple mono-scaling.
However, it preserves the multifractal scaling (\ref{HH1}) in {\it physical} time $T$.

\subsection{Derivation in Terms of Stochastic Processes}

The derivation of the Hurst exponents in the previous section was rather formal.
We now confirm and geometrically explain the results by rederiving them in terms of
stochastic processes $\psi,\psi_n$, which we define as logarithms of time $T$ (\ref{time}) and the moments $m_n$ (\ref{pmom}):
\begin{eqnarray}
T(\hat T)&\equiv&\hat T\cdot e^{\alpha\psi(\hat T)}\ =\  \int_0^{\hat T}d\hat t\ e^{\alpha\tilde\phi(\hat t)},\label{ptime}\\
m_n(\hat T)&\equiv& \hat T^{{n\over2}(1-\Delta_i)}\cdot e^{n\alpha_i\psi_n(\hat T)}\ =\  \int_0^{\hat T}d^n\hat t\ \prod_{k=1}^n e^{\alpha_i\tilde\phi(\hat t_k)}\cdot\langle\prod_{k=1}^n\dot{\hat\pi}_i(\hat t_k)\rangle. \label{pmoment}
\end{eqnarray}
Here, we fix $\phi_{00}=0$, which leaves us with $\tilde\phi$. The effective field $\psi$ would be zero without gravity. 
Given the correlation structure (\ref{fix}) of $\tilde\phi$, the following is shown in the appendix:
\begin{itemize}
\item
The correlation $\rho$ of $\psi(\hat T)$ and the $\psi_n(\hat T)$ is almost 1 for background times $\hat T$ that are much smaller than the correlation time $\tau$.
More precisely, $\rho=1-o(1/\ln [\tau/\hat T])$. 
\item
The variances (connected 2-point functions) of the effective fields are, up to constants:
\begin{eqnarray}
\langle\psi^2\rangle_c&=&\langle\psi_n^2\rangle_c\ =\ (\ln\tau-\ln \hat T)\ \ \ \text{for}\ \ \ \hat T\ll\tau .\label{vari}
\end{eqnarray}
They decrease linearly in background log-time, and become 0 at the correlation time.
\item
The drifts (growth of the expectation values) of the effective fields are, up to constants:
\begin{eqnarray}
\alpha\langle\psi\rangle&=&{\alpha^2\over2}\ln \hat T\ \Rightarrow\ \langle\ln T\rangle\ =\ {Q\alpha\over2}\ln \hat T,\notag\\
\alpha_i\langle\psi_n\rangle\ &=& {\alpha_i^2\over2}\ln\hat T\ \Rightarrow\ \langle\ln M_n\rangle\ =\ n\cdot {Q\alpha_i-1\over2}\ln \hat T \label{drif}
\end{eqnarray}
for $\hat T\ll\tau$, using relations (\ref{DDK}). Thus, the gravitational dressing gives physical log-time 
and the log-moments (\ref{mono}) additional drifts $\alpha^2/2$ and $n\alpha_i^2/2$ in background log-time. 
\end{itemize}

\begin{figure}[t!]\centering
\includegraphics[height=5cm]{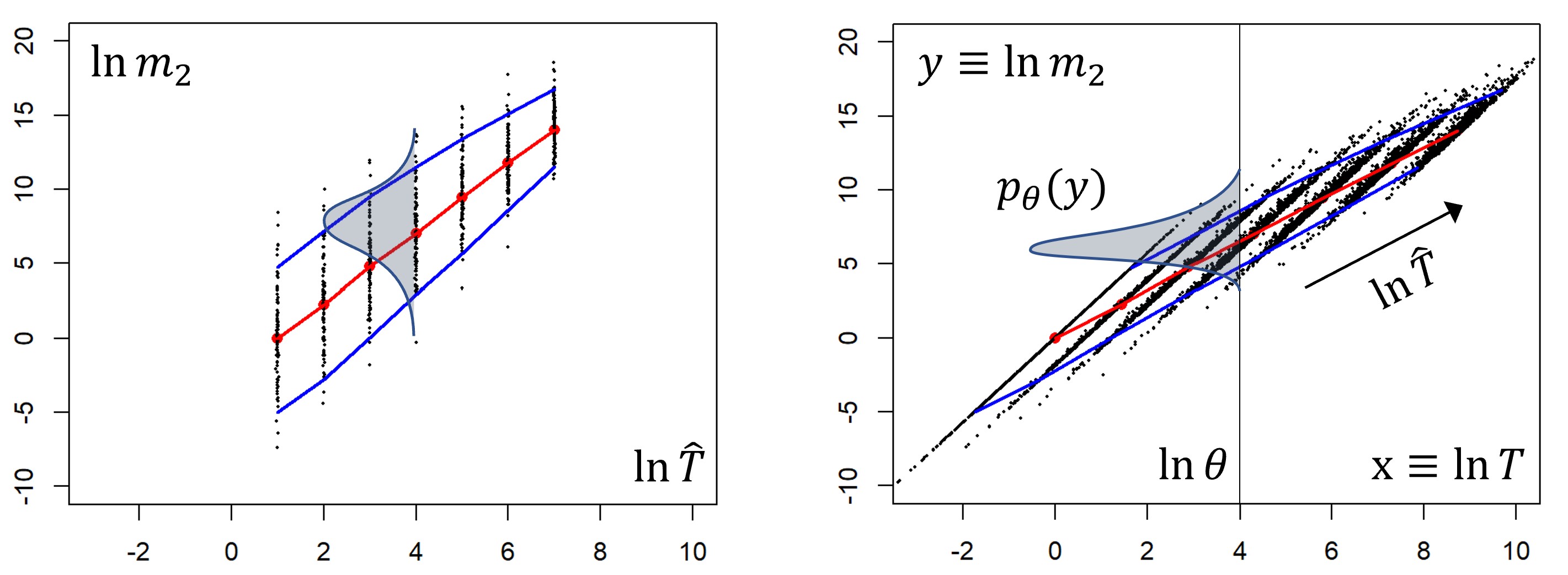}
\caption{Left: $\ln m_2$ as a stochastic process with decreasing variance in background time $\hat T$. Right: scatter plot of $\ln m_2$ vs $\ln T$. 
Their correlation slowly decreases from 1. 
A cross section at fixed physical time $\theta$ yields $\ln m_2$ as a stochastic process in physical time.} 
\end{figure}

Fig. 2 (left) illustrates the situation based on a numerical simulation of $\ln m_2$ for background times $\hat T=0, 4, 16, 64, ...$, measured in units of the time interval cutoff. 
Red and blue lines connect the means and the 5th/95th percentiles of the distribution, which is shown as a gray area.  
The variance of the distribution indeed decreases linearly in log-time (its standard deviation decreases like a square root), while its mean increases linearly.
The scatter plot in fig. 2 (right) shows the {\it bivariate} distributions of $x\equiv\ln T$ vs. $y\equiv\ln m_2$ for the same values of $\hat T$.
The points indeed scatter around parallel, equidistant lines, reflecting the correlation of almost 1. 
For fixed $\hat T$, the distributions of $x$ and $y$ are Gaussian. They are also sketched as dotted lines in fig. 3 (left), approximating the correlation $\rho$ by 1.\\

\begin{figure}[t!]\centering
\includegraphics[height=5cm]{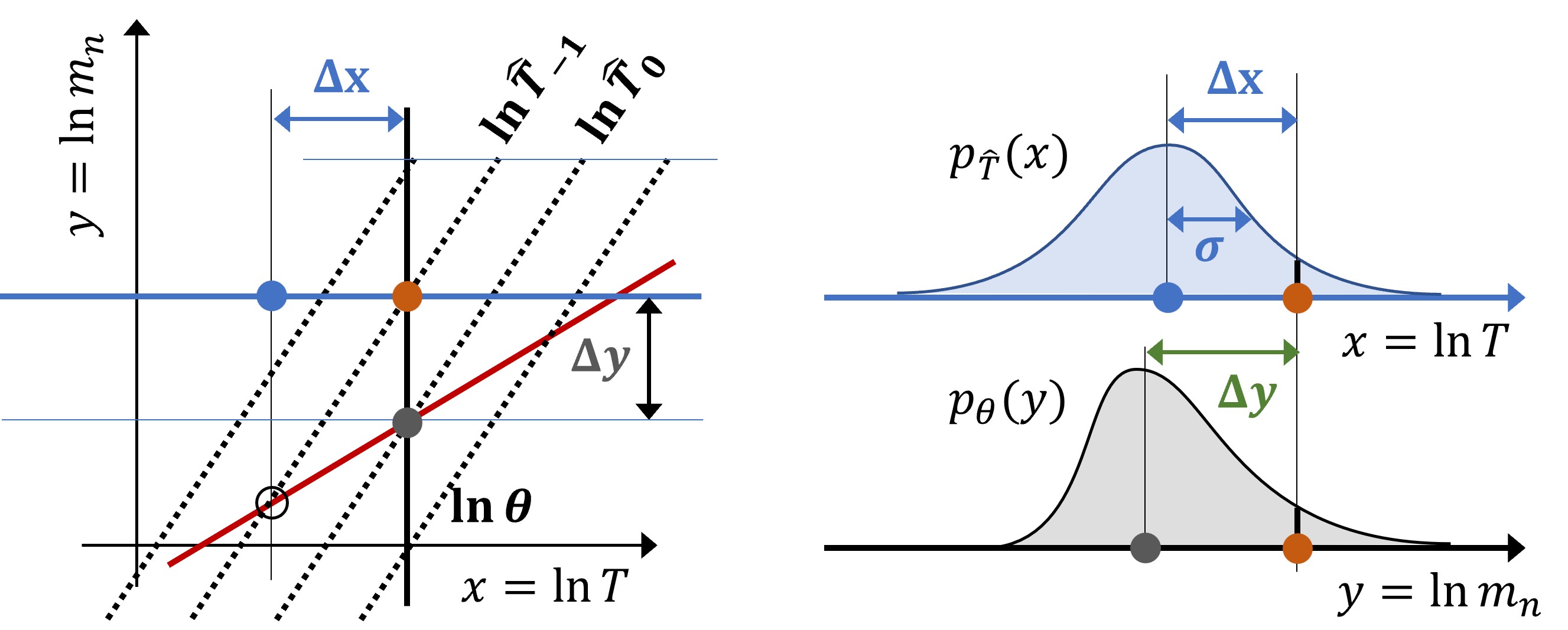}
\caption{Inferring the skewed cross-sectional distribution $p_\theta(y)$ (gray area) along the vertical line of fixed physical time $\theta$ 
from the Gaussian distribution $p_{\hat T}(x)$ (blue area) of physical time for fixed background time $\hat T$. The dotted lines sketch those in the scatter plot in fig. 2 (right).} 
\end{figure}

As explained in subsection 3.4, the moment $M_n$ at physical time $\theta$ is the expectation value of $m_n$, 
restricted to those paths $\tilde\phi(\hat t)$ that lead to the end value $T=\theta$:
$$M_n(\theta)\ =\ \int d(\ln\hat T)\ \Big\langle m_n(\hat T)\cdot\delta\Big(\ln T(\hat T)-\ln\theta\Big)\Big\rangle.$$
Here we must allow all background times $\hat T$.
The delta function cuts the bivariate distribution at a cross section of fixed {\it physical} time $\theta$. This is represented by the vertical lines in fig. 2 (right) and fig. 3 (left).
Let $p_\theta(y)$ be the cross-sectional distribution along these lines (gray area in fig. 2, right, and gray area in fig. 3, bottom right).
$p_\theta(y)$ combines the Gaussian distributions for different, fixed $\hat T$. Since their variance decreases with growing $\hat T$, $p_\theta(y)$ is skewed.
$p_\theta(y)$ represents the stochastic process $\ln m_n(\ln \theta)$ in {\it physical} log-time. 
Its center and its drift in physical time are easily inferred by combining the two equations (\ref{drif}):
\begin{eqnarray}
\langle\ln T\rangle&=& {Q\alpha\over2}\ln \hat T\ \equiv\ \ln \theta,\notag\\
\  \ \ \langle\ln M_n\rangle&=& {n\over2}(Q\alpha_i-1)\ln\hat T\ =\ n({\alpha_i\over\alpha}-{1\over Q\alpha})\cdot\ln \theta.\notag
\end{eqnarray}
The red line in fig. 3 (left) shows the center of the process as a function of physical time.
We can read off the distribution $p_\theta(y)$ from the Gaussian probability distributions $p_{\hat T}(x)$ for fixed $\hat T$ (shown for the example $\hat T_{-1}$ as the blue area in fig. 3, top right)
by matching both at the intersection point of the line of fixed $\theta$ and the line of fixed $\hat T$ (marked as a red dot for the example $\hat T_{-1}$).
Let $\Delta y$ be its distance from the center of the distribution $p_\theta(y)$, 
and let $\Delta x$ be its distance from the center of the Gaussian distribution $p_{\hat T}(x)$. From (\ref{vari},\ref{drif}),
\begin{equation}
\Delta x = {Q\alpha\over2}\Delta\ln\hat T,\ \ \ \Delta y= {n\over2}\Delta\ln\hat T,\ \ \ \sigma=\alpha\sqrt{\ln\hat\tau-\ln \hat T_0+\Delta\ln\hat T},\notag
\end{equation}
where $\Delta\ln\hat T$ is the spacing of background times $ ..., \ln\hat T_{-1}, \ln\hat T_0, ...$ (see fig. 3). This yields
\begin{eqnarray}
p_\theta(y)\sim\exp\{-{\Delta x^2\over2\sigma^2}\}\ =\ \exp\{-V(\Delta y)\}\ \ \ \text{with}\ \ V(\Delta y)\equiv{Q^2\over4n}\cdot{(\Delta y)^2\over nX+\Delta y},\notag
\end{eqnarray}
where $X$ measures the distance from the correlation time: 
$$X\equiv {\ln\tau-\ln\theta\over Q\alpha}={\ln\hat\tau-\ln\hat T\over2}.$$
From the probability distribution, we compute the expectation value of the $n$-th moment:
$$\langle m_n\rangle\ \sim\ \int dy\ p_\theta(y)\ e^{y}\ =\ \theta^{n({\alpha_i\over\alpha}-{1\over Q\alpha})}\cdot \int d(\Delta y)\ e^{-V(\Delta y)+\Delta y}.$$
For large $\tau$, i.e, large $X$, we can approximate the integral by its value at the saddle point:
$${\Delta y\over nX}=\sqrt{Q^2\over Q^2-4n}-1\ \ \Rightarrow\ \ -V(\Delta y)+\Delta y\ =\ -nX\cdot\Big[1-{Q^2\over2n}\big(1-\sqrt{1-{4n\over Q^2}}\big)\Big].$$
This yields the Hurst exponents, which precisely agree with our previous formula (\ref{HH1}).
$$\langle m_n\rangle\sim\theta^{nH_n}\ \ \ \Rightarrow\ \ \  H_n={\alpha_i\over \alpha}-{Q\over2n\alpha}\cdot\big(1-\sqrt{1-{4n\over Q^2}}\big) $$

This second, quite different derivation of the Hurst exponents in terms of stochastic processes not only confirms our results, but also yields some new insights:
\begin{itemize}
\item
Although the noise is distributed normally for fixed background time, its distribution is skewed for fixed physical time.
In this respect it differs from the ordinary MRW. It also differs from the skewed MRW of \cite{bouch6}, which is coupled to the "matter" field $\hat\pi$.
\item
It is clear why the higher moments $M_n$ do not exist for $n\le Q^2/4$: the potential $V(y)$
decays as $\exp\{-y\cdot Q^2/4n\}$ for large $y$, where $y$ is the log of the moments,
so the tails of the distribution of the moments fall off as a power $\vert m_n\vert^{-Q^2/(4n)}$.
Thus the expectation value of $\vert m_n\vert$ diverges for $n>Q^2/4$.
\item
It is interesting that the distribution spreads {\it inversely} with the time scale $\theta$, becoming narrower closer to the correlation time $\tau$. 
In this sense, it seems to dissipate from large to small scales. This is remiscent of turbulent cascades, where large eddies dissipate to smaller eddies,
and where multifractal scaling is also observed. It remains to be seen whether this analogy is coincidental or has a deeper origin. 
\end{itemize}

\section{Examples: Minimal Models}

Let us now illustrate our key results (\ref{HH1}, \ref{HH2})
\begin{eqnarray}
H_n&=&{\alpha_i\over\alpha}-{Q\over 2n\alpha}\Big(1-\sqrt{1-{4n\over Q^2}}\Big) \label{HH}\\
&\approx& {\alpha_i\over\alpha}-{1\over Q\alpha}-{2n\over Q^3\alpha}\ \ \ \ \ \text{for}\ \ Q\rightarrow\infty \notag\end{eqnarray}
at a few examples. This is based on the approximation $z\approx2$, which neglects factors of order $z/2\approx(1+0.7\cdot\Delta)$ that should give only small corrections for small $\Delta$.\\

We consider the unitary and non-unitary minimal models \cite{BPZ}. They are labelled by two co-prime integers $(p,q)$ (we choose $p>q$) and have central charges
$$c=1-{6(p-q)^2\over pq},\ \ p,q\in\{2,3,4,...\}.$$
The unitary minimal models correspond to $m\equiv q=p-1\ge3$. 
For a given model, the primary fields $\Phi_{rs}$ are labelled by two integers $r,s\in N$ (in place of $i$) and have dimensions 
$$\Delta_{rs}={k^2-(p-q)^2\over 2pq}\ \ \ \text{with}\ \ 1\le r\le q-1,\ \ 1\le s\le p-1,\ \ k\equiv pr-qs.$$
We put the minimal models on a genus zero random surface and obtain:
\begin{eqnarray}
\alpha^2&=&{2q\over p}\ ,\ \ \ {Q\over\alpha}=1+{p\over q}\ ,\ \ \ 2{\alpha_{rs}\over\alpha}=1+{p-k\over q}\ ,\ \ Q^2={25-c\over3}=4+2\Big({p\over q}+{q\over p}\Big).\notag
\end{eqnarray}

For the Ising model $(p,q)=(4,3)$ with $c=1/2$ on a random surface with the magnetization $\Phi_{22}$ as an order parameter ($k=2$), there are only two real Hurst exponents (\ref{HH}):
$$\Delta_{22} ={1\over8},\ \alpha^2={3\over2},\ {Q\over\alpha}={7\over3},\ {\alpha_{22}\over\alpha}={5\over6}\ \ \ \Rightarrow\ \ \ H_1={1\over2},\ H_2={1\over3}.$$
As another example, the 3-state Potts model $(p,q)=(6,5)$ with $c=4/5$ has a primary field $\Phi_{2,3}$ (where $\vert k\vert=3$), 
which - if used as an order parameter - yields the parameters
\begin{eqnarray}
\Delta_{23} &=&{2\over15},\ \ \ \alpha^2\ =\ {5\over3},\ \ \ {Q\over\alpha}\ =\ {11\over5},\ \ \ {\alpha_{23}\over\alpha}\ =\ {4\over5}\notag\\
\Rightarrow\ \ \ H_1&=&{1\over10}({\sqrt{61}-3)}\ \approx\ 0.48,\ \ \ H_2\ =\ {3\over10}.
\notag\end{eqnarray}
For the general unitary minimal models with $(p,q)=(m+1,m)$ and $\Phi_{rs}$ as an order parameter, one gets $k\equiv m(r-s)+r$ and obtains the only real Hurst exponents
$$H_1 ={1\over2m}\big(\sqrt{2m^2+2m+1}-k\big) ,\ H_2={1\over2}-{k-1\over2m} . $$
The limit $m\rightarrow\infty$ yields the Hurst exponents for $c=1$ models on a random surface \cite{kleb}:
$$H_1\ \rightarrow\ {1\over\sqrt2}-{\kappa\over2},\ H_2\ \rightarrow\ {1\over2}-{\kappa\over2}\ \ \ \text{as}\ \ \ m\rightarrow\infty\ ,\ \ \ \text{if}\ \ \ {k\over m}\rightarrow{\kappa}\in\ ] 0,1 [\ .$$
As mentioned in section 2, the $c=1$ models also include coset models with global symmetry groups such as $SO(n)$.
We can conserve the corresponding currents by choosing "model J" instead of "model A". This will also change the dynamic critical exponent $z$.
It will be interesting to explore this in the future.\\

An interesting class of {\it non}-unitary minimal models are those with large negative central charge $c\rightarrow\infty$. 
Using the primary field $\Phi_{rs}$ as an order parameter, one obtains from (\ref{HH}):
\begin{eqnarray}
Q^2&=&{25+\vert c\vert\over3}\ \rightarrow\ \infty,\ \ \ \alpha\ \rightarrow\ {2\over Q},\ \ \ {\alpha_{rs}\over\alpha}\ \rightarrow\ 1-{1\over2}\Delta_{rs}\notag\\
\Rightarrow\ H_n&\rightarrow&{1\over2}(1-\Delta_{rs})-{3n\over25+\vert c\vert}.
\notag\end{eqnarray}
Note that the multi-scaling effect disappears in the limit $c\rightarrow\infty$.
Such models can be obtained by setting $p/q\gg1$ so $c\approx -6p/q.$
They can formally be interpreted as $O(n)$ models with $n=-2\cos(\pi\cdot q/p)\approx-2$ in the dense phase \cite{kostau}, and
include the Kazakov models $(p,2)$ and the topological models $(p,1)$. The latter have no bulk fields, but if one allows for surface boundaries,
they have boundary fields. For such models, the random surfaces with boundaries may degenerate to branched polymers \cite{kostau}. \\

As mentioned in the introduction, our results can potentially be applied to financial markets or other systems, for which multifractal scling is observed. This is beyond the scope of this paper.
Let us merely remark that one would need a large negative central charge $c$ of order 100, and $\Delta$ of order $-0.1$ in order to explain the 
Hurst exponents $(H_1,H_2,...)\approx0.52\pm0.03,0.50\pm0.03, ...$ that have been empirically observed in highly liquid financial markets (based on a wide range of estimates including \cite{bouch2,tizi}).
This can only be achieved by non-unitary models with $p/q$ roughly of order 20.
More precise empirical measurements and the search for a model that replicates all stylized facts of finance 
are in progress.

\section*{Acknowledgements} 

I would like to thank Henriette (formerly Wolfgang) Breymann, Uwe Täuber, and Matthias Staudacher for helpful discussions and Jean-Philippe Bouchaud for arising my interest in multifractal scaling.
This research is supported in part by the Swiss National Science Foundation under Practise-to-Science grant no. PT00P2\_206333. \\

\section*{Appendix}

Here, we derive the results quoted in subsection 4.2.
The autocorrelation (\ref{bacry2}) of the Gaussian field $\tilde\phi$ of the MRW in background time $\hat t$ is:
$$\langle\tilde\phi(0)\tilde\phi(\hat t)\rangle=(\ln\tau-\ln\hat t)\ \ \ \Rightarrow  \ \ \ \langle\tilde\phi^2\rangle=\ln\tau$$
with correlation time $\tau$. 
Physical time $T$ in (\ref{ptime}) and the moments $m_n$, viewed as stochastic processes (\ref{pmoment}) are
\begin{eqnarray}
T(\hat T)&\equiv& \hat T\cdot e^{\alpha\psi(\hat T)}\ =\  \int_0^{\hat T}d\hat t\ e^{\alpha\tilde\phi(\hat t)},\label{A1}\\
m_n(\hat T)&\equiv& \hat T^{{n\over2}(1-\Delta_i)}\cdot e^{n\alpha_i\psi_n(\hat T)}\ 
\propto\  \int_0^{\hat T}d^n\hat t\ \prod_{k=1}^n e^{\alpha_i\tilde\phi(\hat t_k)}\cdot\langle\prod_{k=1}^n\dot{\hat\pi}_i(\hat t_k)\rangle. \label{A3}
\end{eqnarray}
To compute the Hurst exponents $H_n$ in physical time $T$ for the process shown in figs. 2,3, we need (i) the drifts, (ii) the variances, 
and (iii) the covariances and correlations of the "effective fields" $\psi, \psi_n$ as functions of $\hat T$. 
Consider the expectation value of $T^k$ for $k\in N$:
$$\hat T^k\cdot \Big\langle e^{k\alpha\psi(\hat T)} \Big\rangle=\Big\langle\big[\int_0^{\hat T} d\hat t\ e^{\alpha\tilde\phi(\hat t)}\big]^k\Big\rangle.$$
The left-hand side can be evaluated by Ito's lemma. The right-hand side can be evaluated by Wick contractions
of $\tilde\phi$ using (\ref{fix}). Combining both yields the scaling:
$$\hat T^k\cdot e^{k\alpha\langle\psi\rangle+{1\over2}k^2\alpha^2\langle \psi^2\rangle_c} 
\sim\  \hat T^k\cdot\Big({\hat T\over \tau}\Big)^{-{1\over2}k(k-1)\alpha^2}\cdot \tau^{{1\over2}k\alpha^2},$$
where $\langle\psi\rangle$ and $\langle \psi^2\rangle_c$ denote, respectively, the drift and variance of the random variable $\psi$.
The last term on the right-hand side (the power of $\tau$) comes from self-contractions of $\tilde\phi$ at the same point in time. Comparing the 
powers of $k$ in the exponents, one concludes:
\begin{equation}
\alpha\langle\psi\rangle\ =\ {\alpha^2\over2}\ln\hat T\ +\ a\ \ ,\ \ \ \langle \psi^2\rangle_c\ =\ (\ln\tau-\ln\hat T)\ +\ b,\label{drivar1}
\end{equation}
where $a$ and $b$ are constants (one can show that $a=0$).  
This yields the drift and the variance of the effective field $\psi$ as claimed in (\ref{vari},\ref{drif}).
We see that the gravitational dressing gives physical log-time an additional drift $\alpha^2/2$ 
in background log-time $\ln \hat T$, and a noise volatility $\alpha$ that shrinks to 0 at the correlation time.\\

The derivation for $\psi_n$ is analogous.
Consider the expectation value of $m_n^k$:
$$\hat T^{{1\over2}nk(1-\Delta_i)}\cdot \Big\langle e^{nk\alpha_i\psi_n(\hat T)} \Big\rangle=
\Big[\int_0^{\hat T}d^n\hat t\ \langle  \prod_{j=1}^n e^{\alpha_i\tilde\phi(\hat t_j)}\rangle\cdot\langle\prod_{j=1}^n\dot{\hat\pi}_i(\hat t_j)\rangle\Big]^k.$$
Again evaluating the left-hand side by Ito's lemma and the right-hand side by Wick contractions, and removing the factor $\hat T^{{1\over2}nk(1-\Delta_i)}$ from both sides yields the scaling:
$$e^{nk\alpha_i\langle\psi_n\rangle+{1\over2}n^2k^2\alpha_i^2\langle \psi_n^2\rangle_c} 
\sim\  \Big({\hat T\over \tau}\Big)^{-{1\over2}nk(nk-1)\alpha_i^2}\cdot \tau^{{1\over2}nk\alpha_i^2},$$
Comparing the powers of $k$ in the exponents, one now concludes:
\begin{equation}
\alpha_i\langle\psi_n\rangle\ =\ {\alpha_i^2\over2}\ln\hat T\ +\ a_n\ \ ,\ \ \ \langle \psi_n^2\rangle_c\ =\ (\ln\tau-\ln\hat T)\ +\ b_n,\label{drivar2}
\end{equation}
where $a_n$ and $b_n$ are new constants. This again confirms (\ref{vari},\ref{drif}).
We observe that the effective field $\psi_n$ has drift $\alpha_i/2$ in log time. The variance of its noise shrinks linearly in log time and becomes zero at the correlation time,
analogously to that of $\psi$.\\

Finally, we derive the covariance and correlation of $\psi$ and $\psi_n$ analogously:
$$\hat T^{1+{n\over2}(1-\Delta_i)}\cdot \Big\langle e^{\alpha\psi(\hat T)+n\alpha_i\psi_n(\hat T)} \Big\rangle=
\int_0^{\hat T}d\hat s\ d^n\hat t\ \langle   e^{\alpha\psi(\hat s)}\prod_{j=1}^n\ e^{\alpha_i\tilde\phi(\hat t_j)}\rangle\cdot\langle\prod_{j=1}^n\dot{\hat\pi}_i(\hat t_j)\rangle.$$
Again evaluating the left-hand side by Ito's lemma and the right-hand side by Wick contractions, and removing the factor $\hat T^{1+{n\over2}(1-\Delta_i)}$ from both sides yields the scaling:
$$e^{\alpha\langle\psi\rangle+n\alpha_i\langle\psi_n\rangle+{1\over2}\alpha^2\langle\psi^2\rangle_c+{1\over2}n^2\alpha_i^2\langle \psi_n^2\rangle_c+n\alpha\alpha_i\langle\psi,\psi_n\rangle_c} 
\sim\  \Big({\hat T\over \tau}\Big)^{-{1\over2}n(n-1)\alpha_i^2-n\alpha\alpha_i}\cdot \tau^{{1\over2}\alpha^2+{1\over2}n\alpha_i^2},$$
Comparing the powers of $k$ in the exponents and using (\ref{drivar1},\ref{drivar2}) yields the covariance
\begin{equation}
\langle\psi,\psi_n\rangle\ =\ (\ln\tau-\ln\hat T)\ +\ c_n,\label{drivar3}
\end{equation}
where $c_n$ is another constant.
The correlation is the covariance divided by the square roots of the two variances:
\begin{eqnarray}
\text{cor}(\psi,\psi_n)&=&  {\langle\psi,\psi_n\rangle_c\over\langle\psi^2\rangle_c^{1/2}\langle\psi_n^2\rangle_c^{1/2}}
 \ =\ {\big[\ln(\tau/\hat T)+c_n\big]\over \big[\ln(\tau/\hat T)+ b\big]^{1/2}\big[\ln(\tau/\hat T)+b_n\big]^{1/2}}\ \sim\ 1-o\Big({1\over\ln \tau/\hat T}\Big)\notag
\end{eqnarray}
(the sign in the last equation must be negative as correlations are $\le1$).
We see that the deviation of the correlation of $\psi$ and $\psi_n$ from 1 is small for $\hat T\ll\tau$. 
This confirms our first claim in subsection 4.2.

\end{document}